\documentclass{iau}
\usepackage{graphicx} 
\usepackage{natbib} 

\def\citeN{\citet}
\def\cite{\citep}

\title[IAUS291.~~The HTRU surveys] 
{The High Time Resolution Universe surveys for pulsars and fast transients} 

\author[M. J. Keith]  
{Michael J. Keith$^1$\\on behalf of the HTRU collaboration}

\affiliation{$^1$CSIRO Astronomy \& Space Science, Australia Telescope National Facility, PO Box 78, Epping, NSW 1710, Australia\\ email: {\tt mkeith@pulsarastronomy.net}}

\pubyear{2012}
\volume{291}  
\jname{\mbox{Neutron Stars and Pulsars: Challenges and Opportunities after 80 years}}
\editors{J. van Leeuwen, ed.} 
\begin{document}

\maketitle

\begin{abstract}
The High Time Resolution Universe survey for pulsars and transients is the first truly all-sky pulsar survey, taking place at the Parkes Radio Telescope in Australia and the Effelsberg Radio Telescope in Germany.
Utilising multibeam receivers with custom built all-digital recorders the survey targets the fastest millisecond pulsars and radio transients on timescales of 64$\,\mu$s to a few seconds.
The new multibeam digital filter-bank system at has a factor of eight improvement in frequency resolution over previous Parkes multibeam surveys, allowing us to probe further into the Galactic plane for short duration signals.
The survey is split into low, mid and high Galactic latitude regions.
The mid-latitude portion of the southern hemisphere survey is now completed, discovering 107 previously unknown pulsars, including 26 millisecond pulsars.
To date, the total number of discoveries in the combined survey is 135 and 29 MSPs
These discoveries include the first magnetar to be discovered by it's radio emission, unusual low-mass binaries, gamma-ray pulsars and pulsars suitable for pulsar timing array experiments.

\end{abstract}


\firstsection 
\section{Introduction}
At the time of writing, the ATNF pulsar catalogue lists more than 2000 known pulsars \cite{mhth05}.
The vast majority of these pulsars were discovered in large radio surveys, and more than half were discovered in surveys with the Parkes Radio Telescope.
Much of this success is due to the low system temperature and large field of view provided by the 21-cm ``Multibeam'' receiver \cite{swb+96}.
This receiver has 13 feeds, allowing us to survey 13 patches of sky at once, drastically reducing the time to survey a large area of sky.
The Parkes Multi-beam Pulsar Survey (PMPS) \cite{mlc+01}, Swinburne Intermediate Latitude Survey \cite{ebvb01} and its extensions \cite{jbo+09} all used this receiver, in total discovering nearly 1000 pulsars.
A 13-beam receiver also requires a `backend' recorder capable of digitising 13 independent signals.
For PMPS and associated surveys, an analogue filter-bank was developed capable of 96 3-MHz frequency channels per beam and recording each channel with 1-bit per 250$\mu$s sample.
Although analogue systems are capable of finer frequency resolution, the complexity and large size of such instruments limited the number of channels.

The 13-beam analogue filter-bank is adequate for the detection of the majority of the pulsar population.
However, for the distinct population of millisecond pulsars (MSPs), which typically have spin periods less than $\sim 30\,$ms, dispersive delays in the interstellar medium can broaden the pulse significantly across the 3~MHz band.
This reduces the sensitivity of the PMPS and similar surveys to MSPs with dispersion measures (DMs) greater than $\sim 25$ cm$^{-3}$pc (c.f. the mean DM of the known pulsar population is 230).
Although harder to detect, MSPs are especially valuable objects.
In the last decade, highlights of MSP science includes the pulsar with the fastest rotation period (in the
globular cluster Terzan 5; \citealp{hrs+06}), the double pulsar system \cite{bdp+03,lbk+04}
and its impressive tests of general relativity \cite{ksm+06}, the `missing link'
between the low-mass X-ray binaries and the MSPs \cite{asr+09}, and detailed study of the interstellar plasma at au scales \cite{yhc+07}.
Furthermore, timing of MSPs provides highly accurate parameters \cite{vbc+09} possibly leading
to the detection of gravitational waves in the near future \cite{hbb+09}.

In addition, 1-bit sampling and limited numbers of channels reduces the capacity to detect and characterise fast transients.
\citet{lbm+07} reported the discovery of a burst of radio emission of unknown origin but seemingly at cosmological distance, which has sparked great debate over the origin of such events.
Although this burst was of great significance, further study of these bursts would be aided by greater dynamic range and higher frequency and time resolution.

\section{BPSR and the HTRU Survey}
\begin{figure}
\begin{center}
  \includegraphics[height=.3\textheight]{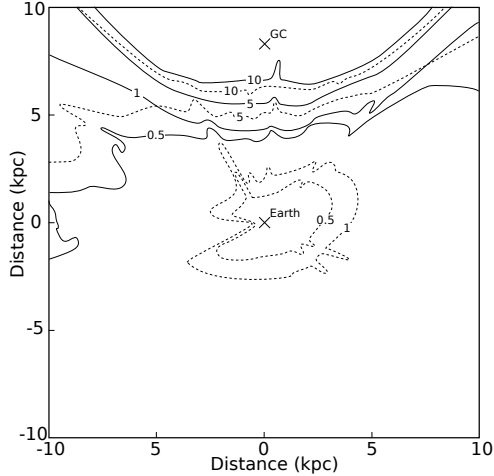}
  \caption{Contours of constant pulse broadening timescale for lines of sight at $b=0$ for a frequency of 1352~MHz.
  The values for the PMPS are shown in dashed lines and for the HTRU survey in solid lines, with contours at 0.5, 1, 5 and 10~ms.
  Scattering timescales and DMs were computed using the NE2001 model \cite{cl02}.
  The Earth and the approximate location of the Galactic Centre (GC) are shown with crosses.
  \label{fig:gal}
  }
  \end{center}
\end{figure}

To improve upon the highly successful Parkes surveys with the 13-beam receiver, we conceived of the High Time Resolution Universe (HTRU) survey for pulsars and fast transients, with the aim of surveying the entire sky with increased sensitivity to MSPs and fast transient events.
This has been made possibly by developing the Berkley-Parkes-Swinburne Recorder (BPSR), an all-digital multibeam pulsar backend.
The digital spectrometers were originally based upon the the Internet Break-Out Board (IBOB) platform developed by the CASPER group at the University of California, Berkley, described in detail in \citet{mcm08}.
In 2012, these were upgraded to use the Re-configurable Open Architecture Computing Hardware (ROACH) platform as part of an upgrade to the backend to support spectral line observations.
The properties of BPSR (as used in the HTRU) and a comparison to those of the analogue system (as used in the PMPS) is given in Table \ref{tab:bpsr}.
For further details on BPSR, see \citet{kjv+10}.
Figure \ref{fig:gal} shows contours of constant pulse broadening time for the PMPS and the HTRU surveys, combining both scattering and dispersion broadening computed from the NE2001 model \cite{cl02}.
Clearly, BPSR provides much greater penetration into the Galactic plane than the filters used in the PMPS, limited only by scattering close to the Galactic Centre.

\begin{table}
\begin{center}
\begin{tabular}{rll}
\hline
  & PMPS & HTRU\\
\hline
Bandwidth (MHz) & 340 & 288 \\
Sample rate ($\mu$s) & 64 & 250 \\
Number of channels & 870 & 96 \\
Channel bandwidth (MHz) & 0.39 & 3 \\
Number of bits per sample & 2 & 1 \\
Data rate (MB~s$^{-1}$) & 52 & 0.62 \\
\hline
\end{tabular}
\caption{Comparison of the BPSR backend with the analogue filter-bank backend as used in the PMPS.}
\label{tab:bpsr}
\end{center}
\end{table}

At Parkes, we are surveying the entire southern sky in three parts:
\begin{itemize}
\item The {\bf Low-latitude} segment covers a thin strip of the Galactic plane, with latitude $|b| < 3.5^\circ$ and longitude $-80^\circ < l < 30^\circ$. The 4300~s integration time is twice that of the PMPS, providing the most thorough survey of the inner Galactic plane to date. Here we primarily target faint pulsars deep in the Galactic plane.
\item The {\bf Mid-latitude} segment covers the majority of the Galaxy, limited by latitude $|b| < 15^\circ$ and longitude $-120^\circ < l < 30^\circ$. The integration time is 540~s, which allows us to cover the large area of sky quickly. Here we focus on bright MSPs suitable for timing array projects.
\item The {\bf High-latitude} segment covers all sky south of declination $+10^\circ$, integrating for 270~s per pointing. Here we primarily target bright MSPs and gather a snapshot of the transient sky at 64~$\mu$s resolution.
\end{itemize}
In addition, we are undertaking a complementary survey of the northern sky with the Effelsberg Radio Telescope at the same sensitivity, using a 7-beam multibeam receiver and digital backend system. More details on the northern HTRU survey can be found in Ng et al. (this proceedings) and references therein.

Observations of the southern HTRU survey started in 2007 and mid-latitude component of the survey is now complete, and observations have been made for $\sim 80\%$ of the low-latitude and $\sim 50\%$ of the high-latitude survey.
Analysis of the survey has so far been limited to solitary pulsars, or those in orbits much longer than the observation time.

\section{Overview of Discoveries}
To date, the combined north and south HTRU surveys have discovered 135 pulsars, of which 29 are MSPs.
Much of the survey is still being analysed, and so here we concentrate on the results of the mid-latitude portion of the southern survey, which accounts for 107 of the total discoveries and 26 MSPs.
This compares favourably with the the predicted discovery rates from \citet{kjv+10}, which indicates that the complete survey could discover as many as 800 previously unknown pulsars and more than 100 MSPs.
Details of the first 12 MSP discoveries appear in \citet{bbb+11b} and \citet{kjb+12}, and the first 75 canonical pulsars in \citet{bbb+12}.
Several of these MSPs may be suitable for pulsar timing array experiments, with typical timing precision less than 1$\,\mu$s.
PSR J1017$-$7156 is especially notable in this regard with root-mean-square of the timing residuals being $\sim0.6\,\mu$s over two years, and this pulsar is already included in the Parkes Pulsar Timing Array.
The rest of this section covers highlights of the HTRU discoveries to date.

\subsection{First Magnetar Discovered by Radio}
The first major result from the HTRU survey was not an MSP, but the discovery of PSR J1622$-$4950, a $4.3\,$s pulsar with a very high inferred magnetic field strength, $\sim 3\times10^{14}\,$G \cite{lbb+10}. The pulsar exhibits large red noise in it's timing residuals and has a highly variable pulse profile.
In addition, the pulsar is clearly detectable up to $24\,$GHz, exhibiting a similar profile morphology to that at lower frequencies \cite{kjlb11}.
These characteristics lead us to conclude that this pulsar is a magnetar, the first to be discovered via its radio emission.
Observations with the Australia Telescope Compact Array show possible evidence of a supernova remnant, and changes in the phase averaged flux density and polarisation \cite{lbb+10,ags+12}.
Further observations at Parkes have demonstrated that although the typical pulse profile is $\sim2\,$s wide, individual profiles are composed of components with widths of $\sim 20\,$ms. The shape of the profile remains relatively constant with frequency, although the individual pulses narrow with frequency~\cite{lbb+12}.
Although the linear and circular polarisation changes in magnitude and by more than $50\%$ from day-to-day, and the position angles swing can change wildly, we find that the linear polarisation is generally consistent with an aligned rotator \cite{lbb+12}.

\subsection{Unusual Low-mass binaries}
Binary pulsars with degenerate companions are broadly split into three groups: intermediate-mass binary pulsars, with heavy white dwarf or neutron star companions (IMBPs; $M_c\sim 1$M$_\odot$); low-mass binary pulsars with light  white dwarf companions (LMBPs; $M_c\sim 0.2$M$_\odot$); and very low-mass binary pulsars that appear to have had a formation process that involves significant mass loss of the companion (VLMBPs; $M_c\sim 0.02$M$_\odot$).
Until recently, it had been thought that VLMBPs underwent a relatively long period of accretion, causing the pulsar to spin up to short spin periods (i.e. $< 10\,$ms) and causing significant mass loss of the companion.
However, the discovery of PSR J1502$-$6752, a $26.7\,$ms pulsar in a 2.48 day orbit around a companion with minimum mass 0.02~M$_\odot$~\cite{kjb+12} challenges that assumption.
The true companion mass does, of course, depend on the inclination angle, and it is possible that we are observing these LMBPs close to face on.
For the companion mass of PSR J1502$-$6752 to be greater than 0.1~M$_\odot$, the inclination angle must be less than $13^\circ$, which has a probability of 0.025 of being drawn from a uniform distribution of three-dimensional orientations.
It is also worth noting that, as mentioned by \citeN{fck+03}, there is a distinct gap in the mass functions of the lowest mass binary MSPs between the VLMBPs.
Therefore, if we disfavour the chance of a face-on system, then either the formation process of VLMBPs does not require a long accretion process, or that PSR J1502$-$6752 has an unlikely formation mechanism and so is inherently rare.
We do not yet understand The formation mechanism of VLMBPs, and the discovery of PSR J1502$-$6752 undoubtedly complicates the picture further.
The discovery of further VLMBPs would be greatly beneficial in determining if PSR J1502$-$6752 really is a unique, or if the spin period distribution of the VLMBPs is in fact wider than currently understood.

The HTRU survey has also uncovered a new class of binary pulsar, the `ultra low-mass binary pulsar' having $M_c\sim 0.001$M$_\odot$, typified by PSR J1719$-$1438 \cite{bbb+11a}. These objects are discussed further in Ng et al. (this proceedings).

\subsection{Polarisation of MSPs}
The improvement in data recording systems and the greatly increasing number of known MSPs has means we now have a much better sample of high quality polarisation measurements of MSPs.
From these observations we see that in many ways the polarisation of MSPs is very similar to the `normal' pulsar population.
In particular, there is a wide variety in the fraction of linear and circular polarisation and a non-negligible fraction are amenable to fitting using a geometric model of the emission, i.e. the rotating vector model (RVM; \citealp{rc69a}).
However, many MSPs have very large pulse widths, especially when observed with high dynamic range~\cite{ymv+11}, and in some cases the polarisation profiles can be very complex and do not obey the RVM~\cite{ovhb04}.
\citeN{kjb+12} argue that this is evidence for a high emission height in MSPs (up to $50\%$ of the light cylinder).
This naturally leads to wide profiles, either because they are `aligned' or have emission from both poles.
In turn, this complicates the profile, making it hard to identify components in the profile and derive emission geometry.
However, in general it seems that the polarised emission of MSPs is not more complex than in many slow pulsars.
There has been success in untangling the complex position angle variations in slow pulsars through studies of polarised intensity on a pulse-by-pulse basis (e.g. \citealp{br80,gl95,kkj+02}).
Future high-sensitivity observations may well be able to repeat this success in MSPs, giving us greater confidence in geometric interpretations of the RVM and a fuller picture MSP emission.
High significance gamma-ray profiles of MSPs could also provide additional constraints on geometry, and test theories of radio emission from the outer magnetosphere.

\subsection{RRATs and Transients}
In addition to the MSP discoveries, we have also discovered 11 Rotating Radio Transients (RRATs; \citealp{bbj+11}).
We have detected many more transient events, with the majority most likely to be RRATs.
We have also detected a number of so-called `pertyons', bursts of terrestrial origin that exhibit frequency dependant delays similar to the $\nu^{-2}$ laws of dispersion in the interstellar plasma \cite{bbe+11}.
However, it is likely that if `extragalactic' bursts such as the event of \citet{lbm+07} are detected in the HTRU surveys, the enhanced time and frequency resolution, in addition to the increased dynamic range, will allow for better characterisation of such bursts.
We have also developed a real-time transient detection system, as part of the upgrade of the BPSR instrument.
This provides important real-time monitoring of RFI, as well as the chance to follow up on significant radio transients within minutes of the burst arriving at the Earth.

\section{The Future of the HTRU surveys}
Although the mid-latitude survey is now complete, we are already re-processing this survey with greater sensitivity to relativistic binary pulsars.
This has the potential to make important new discoveries, including systems more relativistic than the the double pulsar system PSR J0737$-$3039A/B.
In addition, we have barely scratched the surface of the high and low-latitude parts of the southern hemisphere survey, nor much of the northern survey.
The low-latitude survey is the deepest survey of the Galactic plane ever performed, and will make important constraints on the pulsar population, and on the timescales for RRATs and similar nulling pulsars \cite{bbj+11}.
Finally, the high-latitude survey covers a large area of sky that has never before been surveyed at this frequency, and the increased sensitivity to transient events makes will provide a powerful dataset for understanding and characterising short duration radio transients.

\section*{Acknowledgements}
The Parkes radio telescope is part of the Australia Telescope which is funded by the Commonwealth of Australia for operation as a National Facility managed by CSIRO.




\bibliographystyle{mn-cup}

\begin{thebibliography}{}

\bibitem[\protect\citeauthoryear{{Anderson} et~al.}{{Anderson}
  et~al.}{2012}]{ags+12}
{Anderson} G.~E. et~al., 2012\textit{, ApJ}, 751, 53

\bibitem[\protect\citeauthoryear{{Archibald} et~al.}{{Archibald}
  et~al.}{2009}]{asr+09}
{Archibald} A.~M. et~al., 2009\textit{, Science}, 324, 1411

\bibitem[\protect\citeauthoryear{Backer \& Rankin}{Backer \&
  Rankin}{1980}]{br80}
Backer D.~C.,  Rankin J.~M., 1980\textit{, ApJS}, 42, 143

\bibitem[\protect\citeauthoryear{{Bailes} et~al.}{{Bailes}
  et~al.}{2011}]{bbb+11a}
{Bailes} M. et~al., 2011\textit{, Science}, 333, 1717

\bibitem[\protect\citeauthoryear{{Bates} et~al.}{{Bates} et~al.}{2012}]{bbb+12}
{Bates} S.~D. et~al., 2012\textit{, MNRAS}, {in press (arXive:1209.0793)}

\bibitem[\protect\citeauthoryear{{Bates} et~al.}{{Bates}
  et~al.}{2011}]{bbb+11b}
{Bates} S.~D. et~al., 2011\textit{, MNRAS}, 416, 2455

\bibitem[\protect\citeauthoryear{{Burgay} et~al.}{{Burgay}
  et~al.}{2003}]{bdp+03}
{Burgay} M. et~al., 2003\textit{, Nature}, 426, 531

\bibitem[\protect\citeauthoryear{{Burke-Spolaor} et~al.}{{Burke-Spolaor}
  et~al.}{2011a}]{bbe+11}
{Burke-Spolaor} S., {Bailes} M., {Ekers} R., {Macquart} J.-P.,  {Crawford} F.,
  III, 2011a\textit{, ApJ}, 727, 18

\bibitem[\protect\citeauthoryear{{Burke-Spolaor} et~al.}{{Burke-Spolaor}
  et~al.}{2011b}]{bbj+11}
{Burke-Spolaor} S. et~al., 2011b\textit{, MNRAS}, 416, 2465

\bibitem[\protect\citeauthoryear{{Cordes} \& {Lazio}}{{Cordes} \&
  {Lazio}}{2002}]{cl02}
{Cordes} J.~M.,  {Lazio} T.~J.~W., 2002, {NE2001. I. A New Model for the
  Galactic Distribution of Free Electrons and its Fluctuations}, preprint
  (arXiv:astro-ph/0207156)

\bibitem[\protect\citeauthoryear{{Edwards} et~al.}{{Edwards}
  et~al.}{2001}]{ebvb01}
{Edwards} R.~T., {Bailes} M., {van Straten} W.,  {Britton} M.~C., 2001\textit{,
  MNRAS}, 326, 358

\bibitem[\protect\citeauthoryear{{Freire} et~al.}{{Freire}
  et~al.}{2003}]{fck+03}
{Freire} P.~C., {Camilo} F., {Kramer} M., {Lorimer} D.~R., {Lyne} A.~G.,
  {Manchester} R.~N.,  {D'Amico} N., 2003\textit{, MNRAS}, 340, 1359

\bibitem[\protect\citeauthoryear{Gil \& Lyne}{Gil \& Lyne}{1995}]{gl95}
Gil J.~A.,  Lyne A.~G., 1995\textit{, MNRAS}, 276, L55

\bibitem[\protect\citeauthoryear{{Hessels} et~al.}{{Hessels}
  et~al.}{2006}]{hrs+06}
{Hessels} J.~W.~T., {Ransom} S.~M., {Stairs} I.~H., {Freire} P.~C.~C., {Kaspi}
  V.~M.,  {Camilo} F., 2006\textit{, Science}, 311, 1901

\bibitem[\protect\citeauthoryear{{Hobbs} et~al.}{{Hobbs} et~al.}{2009}]{hbb+09}
{Hobbs} G.~B. et~al., 2009\textit{, Publ. Astr. Soc. Aust.}, 26, 103

\bibitem[\protect\citeauthoryear{{Jacoby} et~al.}{{Jacoby}
  et~al.}{2009}]{jbo+09}
{Jacoby} B.~A., {Bailes} M., {Ord} S.~M., {Edwards} R.~T.,  {Kulkarni} S.~R.,
  2009\textit{, ApJ}, 699, 2009

\bibitem[\protect\citeauthoryear{{Karastergiou} et~al.}{{Karastergiou}
  et~al.}{2002}]{kkj+02}
{Karastergiou} A., {Kramer} M., {Johnston} S., {Lyne} A.~G., {Bhat} N.~D.~R.,
  {Gupta} Y., 2002\textit{, A\&A}, 391, 247

\bibitem[\protect\citeauthoryear{{Keith} et~al.}{{Keith} et~al.}{2010}]{kjv+10}
{Keith} M.~J. et~al., 2010\textit{, MNRAS}, 409, 619

\bibitem[\protect\citeauthoryear{{Keith} et~al.}{{Keith} et~al.}{2012}]{kjb+12}
{Keith} M.~J. et~al., 2012\textit{, MNRAS}, 419, 1752

\bibitem[\protect\citeauthoryear{{Keith} et~al.}{{Keith} et~al.}{2011}]{kjlb11}
{Keith} M.~J., {Johnston} S., {Levin} L.,  {Bailes} M., 2011\textit{, MNRAS},
  416, 346

\bibitem[\protect\citeauthoryear{{Kramer} et~al.}{{Kramer}
  et~al.}{2006}]{ksm+06}
{Kramer} M. et~al., 2006\textit{, Science}, 314, 97

\bibitem[\protect\citeauthoryear{{Levin} et~al.}{{Levin} et~al.}{2010}]{lbb+10}
{Levin} L. et~al., 2010\textit{, ApJ}, 721, L33

\bibitem[\protect\citeauthoryear{{Levin} et~al.}{{Levin} et~al.}{2012}]{lbb+12}
{Levin} L. et~al., 2012\textit{, MNRAS}, 422, 2489

\bibitem[\protect\citeauthoryear{{Lorimer} et~al.}{{Lorimer}
  et~al.}{2007}]{lbm+07}
{Lorimer} D.~R., {Bailes} M., {McLaughlin} M.~A., {Narkevic} D.~J.,  {Crawford}
  F., 2007\textit{, Science}, 318, 777

\bibitem[\protect\citeauthoryear{Lyne et~al.}{Lyne et~al.}{2004}]{lbk+04}
Lyne A.~G. et~al., 2004\textit{, Science}, 303, 1153

\bibitem[\protect\citeauthoryear{{Manchester} et~al.}{{Manchester}
  et~al.}{2005}]{mhth05}
{Manchester} R.~N., {Hobbs} G.~B., {Teoh} A.,  {Hobbs} M., 2005\textit{, AJ},
  129, 1993

\bibitem[\protect\citeauthoryear{Manchester et~al.}{Manchester
  et~al.}{2001}]{mlc+01}
Manchester R.~N. et~al., 2001\textit{, MNRAS}, 328, 17

\bibitem[\protect\citeauthoryear{McMahon}{McMahon}{2008}]{mcm08}
McMahon P., 2008, Master's thesis, University of Cape Town

\bibitem[\protect\citeauthoryear{{Ord} et~al.}{{Ord} et~al.}{2004}]{ovhb04}
{Ord} S.~M., {van Straten} W., {Hotan} A.~W.,  {Bailes} M., 2004\textit{,
  MNRAS}, 352, 804

\bibitem[\protect\citeauthoryear{Radhakrishnan \& Cooke}{Radhakrishnan \&
  Cooke}{1969}]{rc69a}
Radhakrishnan V.,  Cooke D.~J., 1969\textit{, Astrophys. Lett.}, 3, 225

\bibitem[\protect\citeauthoryear{Staveley-Smith et~al.}{Staveley-Smith
  et~al.}{1996}]{swb+96}
Staveley-Smith L. et~al., 1996\textit{, Proc. Astr. Soc. Aust.}, 13, 243

\bibitem[\protect\citeauthoryear{{Verbiest} et~al.}{{Verbiest}
  et~al.}{2009}]{vbc+09}
{Verbiest} J.~P.~W. et~al., 2009\textit{, MNRAS}, 400, 951

\bibitem[\protect\citeauthoryear{{Yan} et~al.}{{Yan} et~al.}{2011}]{ymv+11}
{Yan} W.~M. et~al., 2011\textit{, MNRAS}, 467

\bibitem[\protect\citeauthoryear{{You} et~al.}{{You} et~al.}{2007}]{yhc+07}
{You} X.~P. et~al., 2007\textit{, MNRAS}, 378, 493

\end{thebibliography}

\end{document}